\documentclass[11pt,twoside]{article}

\usepackage{asp2006}
\usepackage{epsf}
\usepackage{lscape}

\begin{document}
\def\hi{H{\sc i}}
\title{GHOSTS | Bulges, Halos, and the Resolved Stellar Outskirts of Massive Disk Galaxies}   
\author{Roelof S.\ de Jong and David J.\ Radburn-Smith}   
\affil{STScI, 3700 San Martin Dr., Baltimore, MD 21218, USA}    
\author{Jonathan N. Sick}
\affil{Rice University, Houston, TX 77005, USA}

\begin{abstract} 
  In hierarchical galaxy formation the stellar halos of galaxies are
  formed by the accretion of minor satellites and therefore contain
  valuable information about the (early) assembly process of galaxies.
  Our GHOSTS survey measures the stellar envelope properties of 14
  nearby disk galaxies by imaging their resolved stellar populations
  with HST/ACS\&WFPC2.  Most of the massive galaxies in the sample
  ($V_{\rm rot}$$>$200 km/s) have very extended stellar envelopes with
  $\mu(r)$$\sim$$r^{-2.5}$ power law profiles in the outer
  regions. For these massive galaxies there is some evidence that the
  stellar surface density of the profiles correlates with Hubble type
  and bulge-to-disk ratio, begging the question whether these
  envelopes are more related to bulges than to a Milky Way-type
  stellar halo. Smaller galaxies ($V_{\rm rot}$$\sim$100 km/s) have
  much smaller stellar envelopes, but depending on geometry, they
  could still be more luminous than expected from satellite remnants
  in hierarchical galaxy formation models. Alternatively, they could
  be created by disk heating through the bombardment of small dark
  matter sub-halos.
  We find that galaxies show varying amounts of halo substructure. 
\end{abstract}


Hierarchical galaxy formation in a $\Lambda$CDM cosmology has become
the standard paradigm in recent years. However, our understanding of
the galaxy formation process is incomplete.
Which high redshift galaxy building blocks end up in what kind of
local galaxies? How much of the stellar content of the different
galaxy components (bulge, thin and thick disk, stellar halo) is
created in situ and how much is accreted? How does the current
accretion rate compare to $\Lambda$CDM predictions? To address these
questions we have begun the GHOSTS\footnote{GHOSTS: Galaxy Halos,
Outer disks, Substructure, Thick disks, and Star clusters} Survey,
using HST to perform stellar archaeology in the outskirts of 14 nearby
disk galaxies (8 of which are edge-on).

We obtained HST/ACS observations in the F606W and F814W bands, with
typically 2--3 ACS pointings along the major and minor axes of each
galaxy.  Our observations reach approximately 2 magnitudes below the
tip of the Red Giant Branch (RGB), allowing us to identify distinct
features in a Color-Magnitude Diagram (CMD) that relate to stellar
populations of very different ages \citep[for details
see][]{deJ07}. We can investigate the spatial distribution of stars in
each of these features to constrain formation histories of the
different galaxy components. Using this method we have already
constrained models of disk truncations in NGC\,4244 \citep{deJ07}.

\section{Minor axis surface density profiles}

We select RGB stars from our CMDs and use those to trace the stellar surface density
along the minor axis. RGB stars are ideal as they are abundant in our
CMDs, are indicative of old stellar populations (as expected to be
found in the outskirts of galaxies), and are representative of the
underlying stellar mass. To map the surface brightness profiles in the
central regions of the galaxies we use the integrated light from
Spitzer/IRAC 4.5 micron observations. The Spitzer images provide near
unobscured light profiles, even for our edge-on galaxies. We scale the
RGB surface density star counts such that they match the IR luminosity
profiles in the overlapping region. In this way we derive equivalent
surface brightness profiles directly from the RGB star counts.

\begin{figure}
\mbox{
\epsfclipon
\epsfxsize=0.49\textwidth
\epsfbox[83 207 500 550]{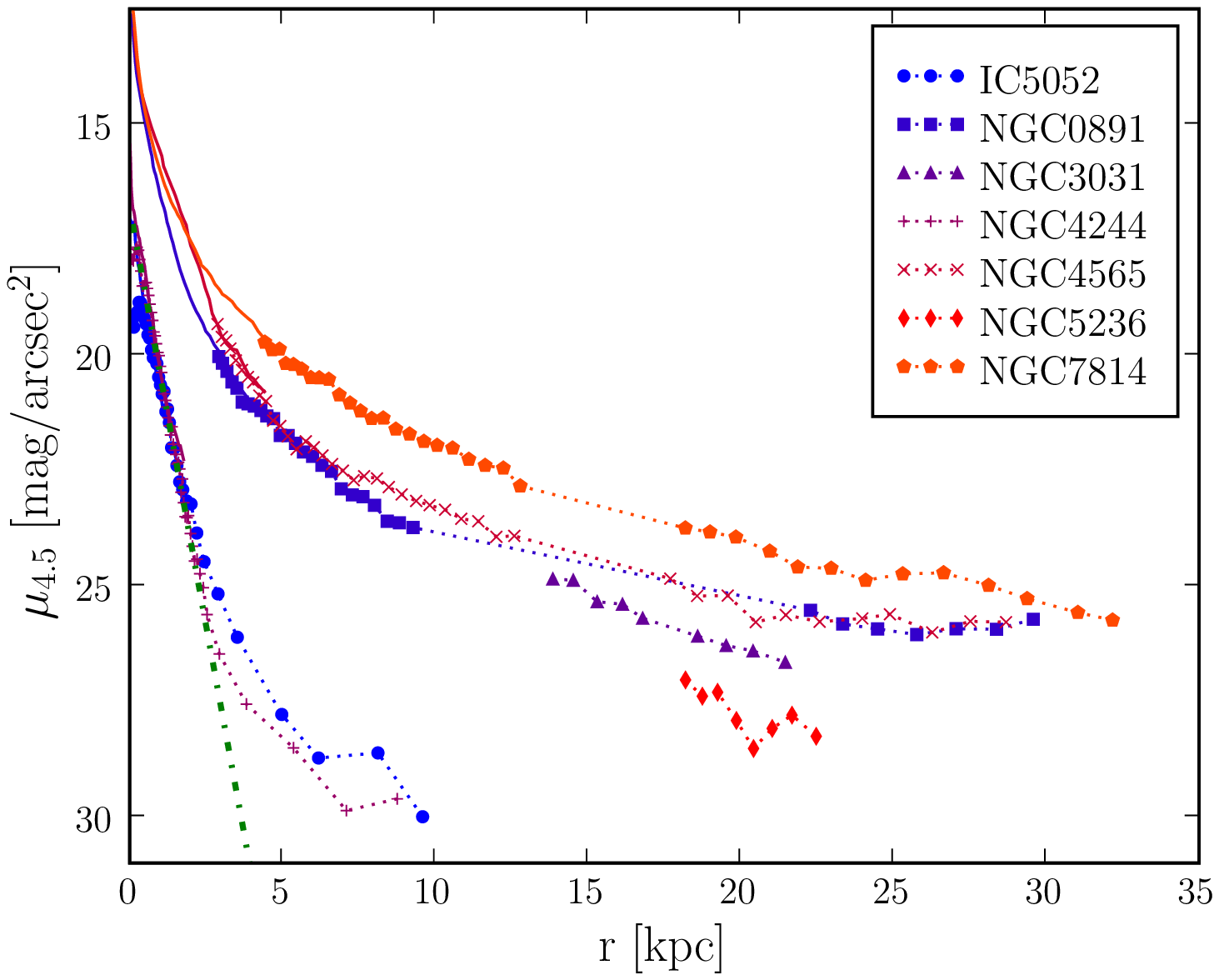}
\epsfclipon
\epsfxsize=0.49\textwidth
\epsfbox[83 207 500 550]{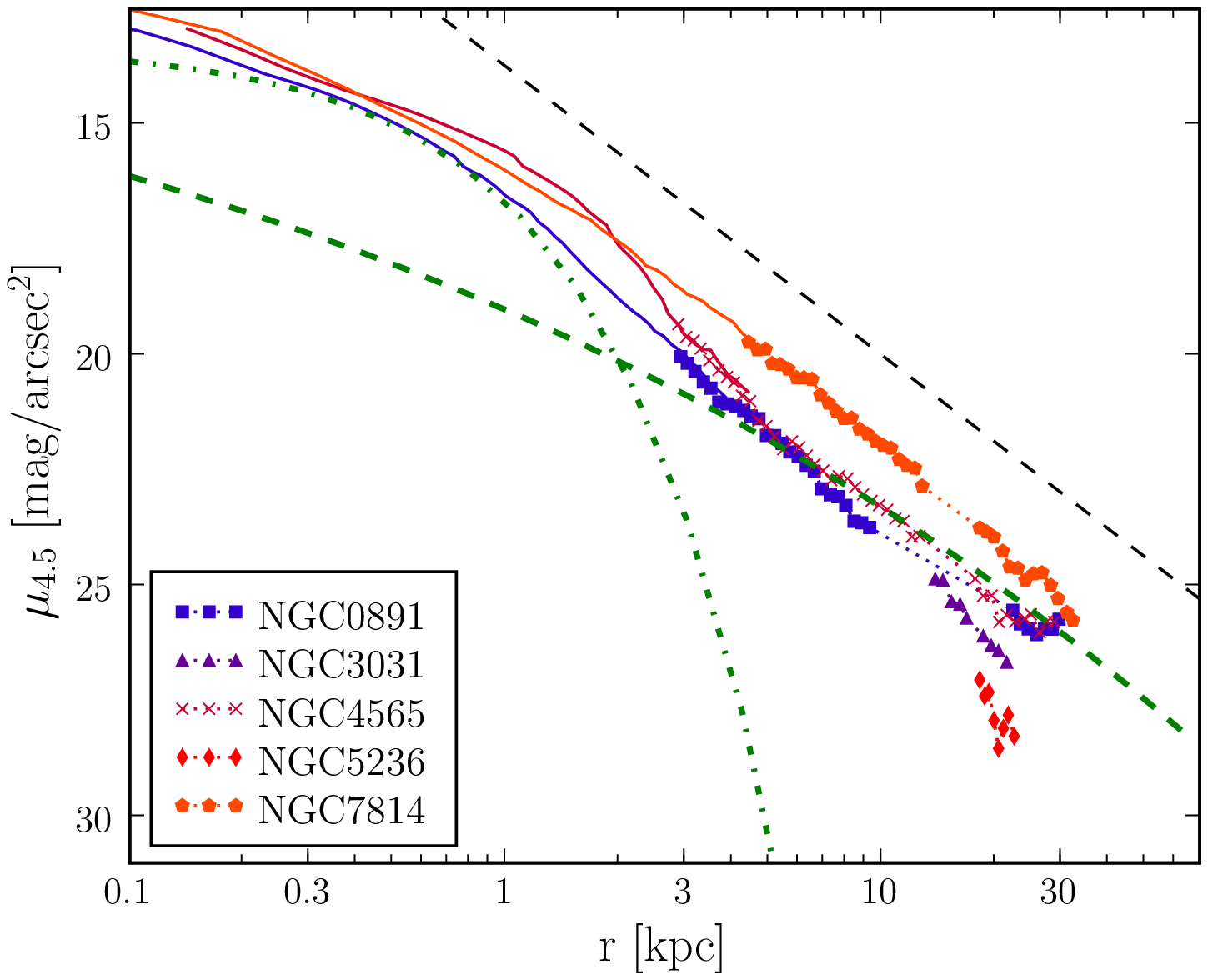}
}
\vspace*{-5mm}
\caption{ Minor axis surface density profiles of GHOSTS galaxies. The
  thin solid lines indicate the profiles derived from Spitzer/IRAC 4.5 micron images
  calibrated to Vega magnitudes (add about 3.5 mag to convert to Vega
  $V$-mag). The symbols connected with dotted lines represent RGB star
  count profiles, scaled to match the Spitzer data. To reduce
  confusion at small radii we only plot star counts beyond 12 kpc for
  the non-edge-on galaxies NGC\,3031/M81 and NGC\,5236/M83. On a
  linear radial scale (left diagram) exponential disks appear as
  straight lines, as indicated for IC5052 by the dot-dashed thick
  line. In the log-log plot on the right, where we have removed low
  mass galaxies for clarity, a straight line indicates a power law
  profile (e.g. thin dashed line = r$^{-2.5}$). Also show are an
  exponential disk (for NGC\,0891, dot-dashed line) and a S\'ersic
  profile with the typical parameters for a flattened stellar halo as
  modeled by \citet{AbaNav06} (thick dashed line).
  \vspace{-1mm}
\label{profs}
}
\end{figure}

In Fig.\,\ref{profs} we show profiles of the edge-on galaxies analyzed
so far, along with outer profiles for the more face-on galaxies
NGC\,3031/M81 and NGC\,5236/M83%
. The exponential thin disks only dominate the inner $\sim$2--3 kpc
(5-10 scale heights), while the extended components are evident at
larger radii.  We find that eight of the nine galaxies analyzed thus
far show components that are more extended than the exponential
disks detected at small radii. Several of the most massive galaxies have very extended
envelopes with stellar densities at 30 kpc that are 10--100 times
higher than the contamination background, equivalent to $\sim$29
$V$-mag arcsec$^{-2}$. NGC\,5236/M83 is the only galaxy with a pure
exponential disk to the last measured point at 20 kpc (more than 10
disk scale lengths).

\subsubsection{The bulge-halo connection}
In this section we explore the connection between bulges and the
extended components. We model the minor axis profiles by combining a
S\'ersic profile and an exponential disk. Merger models show that the
hot components resulting after a violent relaxation generally exhibit
a S\'ersic profile \citep[e.g.,][and reference
therein]{BarHer92,AbaNav06}.  If bulges and stellar envelopes are
created by a collisionless merger processes, we thus expect their
light to follow a S\'ersic profile.

For a number of massive galaxies ($V_{\rm rot}$$>$200 km\,s$^{-1}$) we
can fit the entire minor axis profile over a factor of 1000 in size
($\sim$10$^{4.5}$ in surface density) by an exponential disk and a single
S\'ersic profile, representing both the inner bulge-like region and the
outer envelope. This is, for instance, the case for the bulge
dominated NGC\,7814 or a galaxy like NGC\,3031/M81, which has a power
law outer envelope of rather steep slope. Other galaxies, like
NGC\,4565%
, have too shallow an
outer slope compared to their concentrated bulge region to be fitted
by a S\'ersic profile. NGC\,891 can be fitted by an exponential disk
and a S\'ersic profile from central bulge to outer envelope if we
ignore our outermost field and presume that the higher star density at
30 kpc is due to substructure. Finally NGC\,5236/M83, which has only a
small bulge, shows no sign of an outer envelope out to ten disk scale
lengths.

The smallest galaxies in the sample ($V_{\rm rot}$$\simeq$100--120
km\,s$^{-1}$) have small extended components, barely discernible above
the background contamination (see Fig.\,\ref{profs}). The shape of the
extended component is thus poorly constrained due to both the
uncertainty in the background and low number statistics. The star
counts can be fitted equally well by exponential, power law, and
S\'ersic law profiles. This feature could be the thick disk, as
observations of the NGC\,4244 major axis suggest the component is very
flattened. However, the RGB main disk scale height is already twice
that of the main sequence population and has been argued already
represent the thick disk \citep{Seth05II} with the feature observed
here being an additional component. These additional components are
most likely (depending on exact shape) more luminous than predicted in
the hierarchical models of \citet{PurBul07}, but could have been
created by the bombardment of small dark matter sub-halos
\citep{KazBul07}.

Therefore, a number of the observed extended envelopes seem
structurally directly related to the central bulge regions, like in
NGC\,7814. In other galaxies, where the bulge is too concentrated to
be simply related to the outer envelope, we can suspect that secular
evolution (e.g., bar driven central enhancement and thickening) can
account for the extra (pseudo-)bulge light. NGC\,4565 with its boxy
bulge could nicely fall in this category. In small galaxies
the extended envelope is unrelated to the central region, as these
small galaxies have no bulge.
 
\subsubsection{Envelope properties and halo models}
Comparing the envelopes of the different galaxies we find that the two
small galaxies have much smaller extended components than the larger
galaxies, with surface densities that are lower relative to their
disks. 
The more massive galaxies in our sample are very similar in terms of
mass, luminosity, and scale size. Still, there is significant
variation in outer envelope properties. At 20 kpc NGC\,891, NGC\,4565,
and NGC\,7814 have power law profiles with a slope of about
-2.5. NGC\,3031/M81 has a steeper profile, while at 20 kpc NGC\,5236
is still dominated by the (face-on) disk. At first sight, the envelope
luminosity at 20 kpc seems correlated with Hubble type and
bulge-to-disk ratio, with the bulge dominated NGC\,7814 being the
brightest and the late-type spiral NGC\,5236 showing no sign of an
envelope at all. Although M31 also fits this trend, the Sab galaxy
NGC\,3031/M81 does not, as a steeper and fainter profile is evident at
20 kpc.

In Fig.\,\ref{profs} we also show a typical profile from the
\citet{AbaNav06} model of accreted stars. Our profiles are a bit
shallower and mostly fainter than these models between 10 and 30
kpc. While the surface density normalization may be somewhat uncertain
in the models, the shape is quite well constrained. It could be that
the true halos only dominate at even larger radii and the slope
becomes even shallower at larger radii. However, in hierarchical
galaxy formation the halo and ``classical'' bulge are formed by the
same merging process, so there is no reason to suspect a large
structural change between bulge and halo. The S\'ersic radii derived
for our combined envelope and bulge fits are typically eight times
smaller than those of \citet{AbaNav06}. However, a number of
simulation parameters affect the concentration of the accreted
halos. Increasing star formation suppression in small sub-halos, such
that only the most massive dark matter sub-halos contain stars, yields
steeper and fainter envelopes \citep{BekChi05}. Alternatively, the
stars in the accreted satellites could sit deeper in the potential
wells of their dark matter sub-halos than simulated in these models,
thereby only being tidally stripped closer to the main galaxy, also
resulting in more concentrated halos \citep{BulJoh05}.

\section{Halo substructure}


We use Delaunay Tessellations to reconstruct the surface density
distributions in our fields (Sick \& de Jong, in prep.). NGC\,4631 has
an obvious overdensity associated with the neighboring galaxy
NGC\,4627 just to the north, but there is also a clear overdensity to
the northwest. This overdensity, that is seen in main sequence, AGB,
and RGB stars, is potentially associated with an \hi\ stream
in this strongly interacting system.
In stark contrast to these clear signs of substructure, we find no
 signs of any substructure surrounding the disk of NGC\,4565. When
comparing these density reconstructions with the substructure observed around M31 on the
same physical scale, the NGC\,4565 halo appears to be much
smoother. We are currently developing techniques to quantify halo
substructure.

%


%

\acknowledgements 
Support for Proposal numbers 9765, 10523, and 10889 was provided by NASA
through a grant from the Space Telescope Science Institute, which is
operated by the Association of Universities for Research in Astronomy,
Incorporated, under NASA contract NAS5-26555.


\end{document}